\documentclass[pdflatex,sn-mathphys-num]{sn-jnl}

\usepackage{graphicx}%
\usepackage{multirow}%
\usepackage{amsmath,amssymb,amsfonts}%
\usepackage{amsthm}%
\usepackage{mathrsfs}%
\usepackage[title]{appendix}%
\usepackage{xcolor}%
\usepackage{textcomp}%
\usepackage{manyfoot}%
\usepackage{booktabs}%
\usepackage{algorithm}%
\usepackage{algorithmicx}%
\usepackage{algpseudocode}%
\usepackage{listings}%
\usepackage{hhline}
\usepackage{array}

\theoremstyle{thmstyleone}%
\theoremstyle{thmstyletwo}%

\theoremstyle{thmstylethree}%

\begin{document}

\title[Article Title]{Retrieval, not hallucinations, will be the limiting factor for LLM-based clinical AI tools}


\author*[1]{\fnm{Kirk} \sur{Roberts}}\email{kirk.roberts@uth.tmc.edu}
\author[2]{\fnm{Steven} \sur{Bedrick}}\email{bedricks@ohsu.edu}
\author[3]{\fnm{Kurt} \sur{Miller}}\email{miller.kurt@mayo.edu}
\author[2]{\fnm{William R.} \sur{Hersh}}\email{hersh@ohsu.edu}
\author[4]{\fnm{Hongfang} \sur{Liu}}\email{hongfang.liu@austin.utexas.edu}

\affil[1]{\orgdiv{McWilliams School of Biomedical Informatics}, \orgname{The University of Texas Health Science Center at Houston}, \orgaddress{\city{Houston}, \state{TX}, \country{USA}}}

\affil[2]{\orgdiv{Division of Informatics, Clinical Epidemiology and Translational Data Science}, \orgname{Oregon Health \& Science University}, \orgaddress{\city{Portland}, \state{OR}, \country{USA}}}

\affil[3]{\orgdiv{Center for Digital Health}, \orgname{Mayo Clinic}, \orgaddress{\city{Rochester}, \state{MN}, \country{USA}}}

\affil[4]{\orgdiv{Dell Medical School}, \orgname{The University of Texas at Austin}, \orgaddress{\city{Austin}, \state{TX}, \country{USA}}}



\abstract{Discussions around large language model (LLM) errors in clinical artificial intelligence (AI) generally center around precision errors like hallucinations. This perspective, targeting both clinicians and AI researchers, seeks to shift that discussion to recall errors, particularly in retrieval of patient-level data needed for many clinical AI tools. The perspective outlines types of errors and mitigation strategies, describes research directions in LLMs and retrieval, and provides an overview of retrieval evaluation.}




\maketitle

\section{How patient-level retrieval interacts with LLMs}\label{sec1}


Artificial intelligence (AI) tools, especially those based on large language models (LLMs), have already made sizable impacts on the practice of clinical medicine \cite{Chen2026.nat_med}.
These include tools such as ambient AI scribes that create a draft clinical note from a recorded patient-provider conversation as well as problem-specific patient summarization tools to give providers a quick overview of a patient’s medical history in a problem-specific manner.
Nearly all these tools either require or stand to benefit from the incorporation of the patient’s medical history from the electronic health record (EHR).
This paper thus focuses on errors related to retrieving relevant EHR information from a patient’s history to support LLM-based clinical AI tools.

\begin{figure}
\centering
\includegraphics[width=0.9\textwidth]{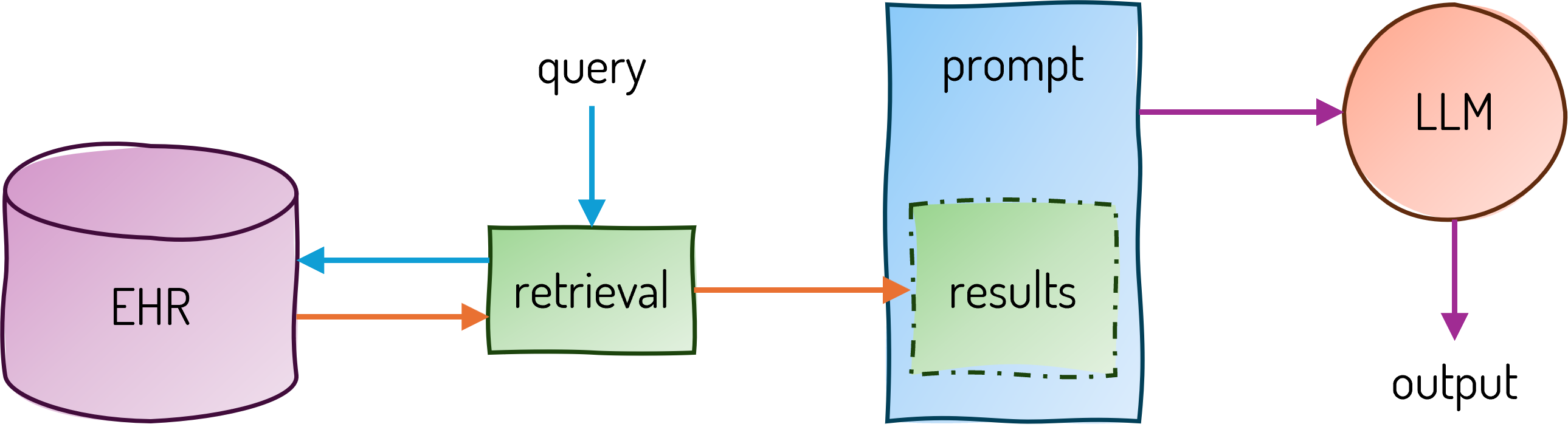}
\caption{Generalized architecture of LLM-based systems incorporating patient-level retrieval. A query (e.g., a medical problem) is given to a retrieval engine; the retrieval engine searches for relevant patient history data from the EHR (e.g., notes/paragraphs that describe the medical problem); the search results are included in a prompt, which is provided to the LLM; the LLM generates some output.}
\label{figure:rag}
\end{figure}

Incorporating the patient’s history into LLM-based tools faces a substantial limitation with regards to using the entirety of a patient’s (often very long, and with a complex structure) record in a single LLM call.
This is often impossible due to the LLM’s context length, but regardless it is generally undesirable to include tremendous amounts of irrelevant information in an LLM prompt and it is certainly costly and slow to utilize so many unnecessary tokens. 
The generally-accepted solution to this is to perform an initial retrieval (search) step to identify the portions of a patient’s record that are relevant for a given task.
This can be seen in Figure~\ref{figure:rag}.
The use of retrieval to provide an LLM with external information (e.g., from the EHR or other sources like the scientific literature) is often referred to as retrieval-augmented generation (RAG).
See Table~\ref{table:examples} for example clinical AI applications involving retrieval.
As stated above, a RAG-based approach is either required for clinical AI tools (e.g., summarizing, diagnosing, or answering questions about patient data) or would substantially benefit them (e.g., providing context to AI scribes to disambiguate or clarify aspects of a patient-provider conversation).
It is thus critical to discuss the types of errors that can be made by an AI tool with a retrieval component and how these errors can be mitigated (or not) in end user applications.

\begin{table}[t]
    \centering
    \begin{tabular}{p{0.8in}|>{\raggedright}p{0.8in}|>{\raggedright}p{0.8in}|>{\raggedright}p{0.8in}|>{\raggedright}p{0.8in}|}
\hhline{~----}
\bf{Task}
  & Diagnosis \hbox{recommendation}
  & Problem-oriented  \hbox{summarization}
  & Question  \hbox{answering}
  & Note \hbox{generation} \\
\hhline{~----}
\bf{Brief \hbox{description}}
  & Given patient signs \& \hbox{symptoms} and access to patient's medical history, recommend appropriate diagnoses for consideration
  & For each patient diagnosis, provide summary of patient's medical history 
  & Given natural language question about patient's medical history, return concise answer
  & Drafting a clinical note (e.g., using ambient scribing) augmented with data extracted from patient record \\
\hhline{~----}
\bf{Query}
  & medical concept (likely agent-driven)
  & problem
  & question
  & open-ended \\
\hhline{~----}
\bf{Precision concerns}
  & Spurious patient history in support of incorrect diagnosis
  & False information not part of patient's history
  & Answer based on incorrect information
  & Generated note contains hallucinations or other false information about patient \\
\hhline{~----}
\bf{Recall \hbox{concerns}}
  & Missed critical patient history needed to justify correct diagnosis
  & Omitted information from history of specific problem
  & Non-answer or answer based on outdated information
  & Generated note is missing critical patient history information \\
\hhline{~----}
\bf{\# of retrieval calls}
  & Open-ended (especially with agent-based methods)
  & One or more per problem
  & One or more per problem
  & Open-ended (likely proportional to note length, unit length, or patient completion) \\
\hhline{~----}
    \end{tabular}
    \caption{Example LLM-based clinical AI tasks with corresponding concerns related to precision and recall.
    Note that different implementations of such systems may work in substantially different ways.
    This table just provides brief examples for consideration of the impact of precision and recall errors.}
    \label{table:examples}
\end{table}

\section{Errors in precision \& recall: their solutions \& consequences}\label{sec2}



There are numerous taxonomies for classifying the risks of using LLMs and the particular types of errors they make (e.g., \cite{Asgari2025}). 
Since our focus here is limited to patient-related clinical AI systems (i.e., systems that process a specific patient's information), it is helpful to draw on the analogy of type I (false positives) and type II (false negative) errors.
Type I errors impact a system’s precision (positive predictive value), giving answers that are not justified by the data, while type II errors impact a system’s recall (sensitivity), resulting in answers that fail to take into account information that would classify that sample as positive.
The type I vs. II error categories do not perfectly map onto the output of a generative AI system: they are precise, discrete categories of errors whereas the output of an LLM is often a paragraph-length response that can be wrong (or simply sub-optimal) in multifactorial ways.
But it is useful to map the types of errors that LLMs make onto the type I vs. II framework because it helps to elucidate the source of (and therefore solution to) certain types of errors.
Thus, in this work we consider errors of precision (type I errors) to be cases in which the LLM output contains claims that are not substantiated by the underlying patient’s data, whereas we consider errors in recall (type II errors) to be cases in which the LLM output is missing important information that is substantiated in the patient’s complete data.
Of course, a given LLM output could contain {\it both} types of errors, as well as multiple instances of either or both.
But here we consider each individual error in precision or recall and probe its potential source.


Precision (type I, or positive predictive value) errors (or ``false alarms'') in this context involve the LLM making some claim that is {\it not} substantiated in the patient’s data.
This includes the class of errors commonly referred to as ``hallucinations'', though there are others.
The term hallucinations is commonly defined as the presence of inaccurate information in the output of an LLM system.
Smith et al. \cite{Smith2023.plos_dig_health} argue that the term hallucination, borrowed from psychiatry, misrepresents what is happening within the LLM inference process and that ``confabulation''--defined as ``the generation of narrative details that, while incorrect, are not recognized as such''--is instead more indicative of these classes of LLM errors.
Regardless, we argue that the inclusion of any information in the LLM output that is not substantiated by the patient’s data can be considered a false positive, and thus an error in precision.
Note that there are other errors in precision that an LLM can make: the terms hallucination/confabulation generally only apply to the natural language outputs of an LLM (e.g., a generated paragraph), but not structured outputs (e.g., when the LLM is prompted to output in JSON).
In the case of a structured output, this would be considered an error in precision when the LLM makes a prediction (e.g., the patient has a particular phenotype) that is not substantiated by the patient’s underlying data.


Errors that affect precision will, at best, waste user time (e.g., reading past irrelevant information) and, at worst, lead the user to an incorrect understanding (e.g., with hallucinations).
Precision errors that lead to decreased human efficiency are not trivial since most currently implemented clinical AI systems exist to save human effort and reduce burden, but exactly because this is their main goal we do not generally see clinician complaints of clinical AI tools wasting their time with excessive output.
Instead, precision concerns generally focus on hallucinations.
While there are many LLM techniques focused on reducing hallucinations \cite{Maynez2020.acl}, the generally-accepted mitigation strategy is to link to the source information (e.g., clinical notes) from which a particular claim is derived. 
While not a perfect solution (e.g., due to low verification rates), the wide implementation of such reference linking provides a clear path to properly vetting the output of a generative AI tool.
Most critically, the other type of error AI systems make lacks a comparable vetting capability.



Recall (type II, or sensitivity) errors (or ``omissions'') in this context involve the LLM {\it not} making some claim that is substantiated in the patient’s data.
In other words, important information about the patient is missing from the LLM’s output.
This could be missing an important diagnosis or medication, or it could mean missing the most recent information (e.g., basing a decision on an out-of-date lab result). 
One could measure recall errors at the document level (e.g., which clinical notes should have been retrieved?) or at the aspect/instance level (e.g., what discrete medical facts should have been retrieved? even if some notes containing that fact were missed) \cite{Hersh2001.ipm}.
The latter is more medically informative, while the former is preferred by information retrieval researchers.
Regardless, since most retrieval methods model query-document similarity, missing any relevant document runs the risk of missing important facts from a patient’s history. 
From Figure~\ref{figure:rag}, we can derive two main fault sources for missing information in LLM output: first, a scenario in which the retrieval system itself (within the RAG approach) failed to return important information, and second, a scenario in which the retrieval system returned complete information, but the LLM failed to include it in the final output.
Regardless of the component that caused the recall error, the AI output will result in clinical decisions being made without critical information about the patient.


While precision errors such as hallucinations can be mitigated with referencing source material, there is no equivalent mitigation strategy for recall errors.
The only complete solution is the traditional process of reviewing the patient’s full record, but this defeats the purpose of recent clinical AI tools (i.e., saving clinicians’ time and assisting with information overload).
Later we discuss how retrieval methods have traditionally been evaluated, but for here it is sufficient to state that, generally speaking, on sufficiently large datasets it is impossible to know the exact recall performance of a system, only a rough upper bound estimate of its performance.
This means that for clinical AI tools that retrieve patient-level information from a patient with a substantial amount of medical data, recall errors are likely to go undetected.
Recall errors may quite literally be a silent killer.


The primary motivation for this perspective article is to point out the contrast between the above point (recall errors are far harder to mitigate than precision errors) and the current discussion around the types of errors that LLM-based clinical AI tools make, which focus almost entirely on hallucination.
For example, in an ongoing qualitative study being conducted by the lead author, all 8 of the clinicians interviewed thus far brought up hallucinations as a concern. 
None, however, had given serious thought to recall errors and had only even considered them after prompted by the interviewer.
Yet after that prompting, all agreed they pose a significant concern.
We grant that hallucinations are a disturbing problem. We also acknowledge that we are not aware of any large-scale study assessing the impact of recall errors in clinical AI tools—but such a study is difficult to conduct precisely for the reasons discussed above: measuring recall is hard.

\section{Advances in LLMs vs. retrieval}


Predicting the future is for clinical AI is fraught with challenges, but we can look at the (recent) past to draw some reasonable inferences on where things are headed. 
Transformer-based language models have made profound levels of progress since their introduction in 2017 \cite{Vaswani2017}, to the extent that the moniker ``large language model'' has been needed to differentiate the later transformer models from those coming earlier.
Even since the widespread use of the term LLM, they have continued to improve at an impressive pace \cite{Zhao2026.arxiv.survey_llm}.

Meanwhile, the pace of improvement in core information retrieval is relatively minimal.
To be clear, there has been significant overall improvements in retrieval as a result of LLMs and other machine learning methods \cite{Zhu2025.acm_tis}, but these improvements largely wrap the core retrieval approach (e.g., as a ``re-ranker''). 
The core retrieval engine typically uses BM25 \cite{Robertson2009.ftir} (or similar TF-IDF variants) or more recent ``dense'' vector-based approaches using embedding models \cite{Zhao2024.acm_tis} that themselves can be derived from LLMs or other transformers.
Importantly, dense models often require interpolation or other modifications to perform as robustly as BM25 \cite{Wang2021.ictir}.
As one example of the robustness issue, the first round of the TREC-COVID challenge in 2020 was won by a basic retrieval system from the 1960s \cite{Salton1965,Buckley1985}, but in the third round (by which time data was available to tune systems) more modern retrieval techniques were clearly superior \cite{Roberts2021.jbi}.

It is likely that continued improvements in dense retrieval will directly follow the improved LLM-adjacent embedding models, but the core focus in research is on the LLM models themselves.
In other words, performance advances in retrieval will likely lag behind the corresponding performance advances in LLM-based methods.
A corollary to this is that the types of LLM precision errors discussed above are far more likely to be solved than the recall errors coming from core retrieval false negatives.
We should not, therefore, assume recall errors will naturally be solved in the course of other machine learning improvements.
Hence the claim we make in the title of this paper: retrieval-related recall errors will increasingly be the limiting factor of clinical AI tools, not precision concerns such as hallucinations.

\section{How retrieval is (and could be) evaluated}


Some insight into the challenges of advancing methods in retrieval for clinical AI can be gleaned from the process of building a retrieval evaluation benchmark.
By comparison, benchmarks in text generation tasks (e.g., scribing, summarization) can be created from existing data (e.g., an existing clinical note) or by creating expert-authored responses that can be compared to the AI-generated text using metrics such as ROUGE \cite{lin2004.text_sum} and BLEU \cite{papineni2002.acl}.
While generation tasks are fundamentally challenging to evaluate for semantic equivalence \cite{Howcroft2020,Dudy2021,Mayfield_2024}, the creation of benchmark data itself is quite feasible.
Another language-related AI task, information extraction (e.g., named entity recognition, relation identification), uses benchmarks where experts manually label data that can be compared to AI predictions using metrics like AUROC or F1.
In contrast to generation and extraction, information retrieval is far more precarious to benchmark.
The best practice retrieval evaluation has been developed over the course of decades by NIST in the TREC evaluations \cite{Voohees2005.book}.
Because the space of possible items to retrieve (the ``document collection'') contains far more items than can feasibly be manually labeled for a given query, the manual labeling must be performed on a relatively small subset.
The best practice method for identifying this subset is a to run a diverse group of actual retrieval systems and include the top $k$ ranked results from each system in the subset (a process known as ``pooling'').

A common concern for retrieval evaluation is that the assessment of an item (its ``relevance'' to the query) is typically seen as more subjective than other language assessments. 
However, Voorhees \cite{Voorhees1998.sigir} showed that while relevance subjectivity affects a system’s score metrics, it has relatively minimal impact on the relative ranking of retrieval systems (e.g., system A and system B are unlikely to flip in relative superiority if a different person performs relevance judging).
A more critical concern, however, is the importance of system diversity within the pool. Systems that retrieve very different items than those used to build the pool are at a disadvantage (such as what happened when the initial dense retrieval systems were compared to older retrieval systems in TREC-COVID \cite{Roberts2021.jbi}). This is, again, a recall issue: we do not know when relevant results are missing from the benchmark. Retrieval recall is simply fundamentally difficult to evaluate.

A somewhat different process is used to evaluate systems that generate responses like a summary of the clinical notes in an electronic health record. In contrast to a retrieval system that produces answers from a fixed corpus, generative systems can produce an unlimited number of different outputs. The BLEU and ROUGE metrics mentioned above compute the overlap of words or sequences of words between the generated text and one or more manually-written examples. These methods worked well in the past when such systems were essentially cutting and pasting existing text, but they don't work for current systems that freely generate text based on their training data. One current evaluation method is based on information ``nuggets'': a human evaluator creates a set of atomic pieces of information that a perfect generated output should contain, and an assessor identifies which of those nuggets are expressed in the generated output \cite{nenkova-passonneau-2004-evaluating, lin-demner-fushman-2006-will}. Nuggets can be concepts, facts, or even questions with answers, but they need to be sufficiently fine-grained that systems only partially address a nugget. When the generated output is required to cite its sources (or provide pointers into the health record), a more thorough evaluation can be done where the sources are checked to ensure they exist, are relevant, and support the citing sentence \cite{Mayfield_2024}. Similar concerns regarding subjectivity and coverage exist for these methods as well, and measuring their impact is the subject of current research. The critical advantage for nugget evaluation is that it can measure recall as well as precision, which BLEU and ROUGE cannot.


While there is promise for using LLMs themselves to evaluate retrieval and generation (``LLM-as-a-judge'') \cite{Dietz25.ictir}, this offers only an approximate expedient. LLMs cannot be a gold standard to judge retrieval because, as Soboroff points out, ``retrieval and evaluation are the same problem'' \cite{Soboroff_2025}. If the LLM were capable enough to grade output reliably, it could also be the system that produced the output. Who measures the grader? As the same time, the pace of progress in clinical AI methods makes the TREC-style benchmark process highly burdensome. And retrieval benchmarks on the few publicly available EHR datasets (e.g., MIMIC \cite{Johnson2016.sci_data,Johnson2023.sci_data}) may well fail to generalize to the EHR data at many institutions. It would be not merely burdensome, but completely infeasible for every institution to perform a TREC-style evaluation each time they intend to implement a clinical AI tool that utilizes a patient-level retrieval component.

To avoid the trap of recall errors silently leading to patient harm, it is likely that some form of hybrid retrieval evaluation should be performed as part of a clinical AI tool implementation. As Soboroff makes clear, some amount of manual assessment is necessary. Adding automation to nugget evaluation processes is a subject of active research \cite{farzi_dietz_2024, pradeep_2024, walden2026autoarguellmbasedreportgeneration}. The great hope in the LLM-as-a-judge paradigm is to reduce the need for that manual assessment, not eliminate it. Research in EHR-based retrieval should focus on how to identify the appropriate mixture of human and LLM-based assessment so as to minimize recall errors while maintaining the feasibility of AI implementations.

\section{Key takeaways}

The purpose of this perspective is to raise awareness of the issue of retrieval recall for clinical AI tools, which increasingly rely on retrieval as part of their core pipeline.
Evaluation of the output of these LLM-based systems can generally only identify precision errors, not recall errors caused by either the LLM not being exposed to important patient information via RAG or not including that important information in its output.
There is no general mitigation strategy for identifying recall errors that also conforms to the productivity goals of the AI tool.

Our point is not to dramatically shift research away from LLMs and hallucinations. We note that some improvements in LLMs will allow for a reduction in recall errors (e.g., improving LLM context length will lead to RAG approaches that provide more patient information to the model).
However, as EHR data continues to increase for patients (e.g., via interoperability), the amount of patient data that retrieval systems search over will increase as well. 
This problem will be particularly pernicious for tasks requiring searches deep into a patient’s medical history.

We assume that the rapid pace in clinical AI will continue.
Research to reduce and mitigate precision issues such as hallucination should continue as well.
Research into recall issues, however, should be amplified, in particular the evaluation of recall errors of a given clinical AI tool in order to assess its potential impact on clinical decision-making and patient health.

\section{Acknowledgements}

This work was partially supported by grants R01LM011934 and R01LM014508 from the National Library of Medicine, NIH.


\bibliography{sn-bibliography}

\end{document}